# Extraordinary Phase Coherence Length in Epitaxial Halide Perovskites

Kostyantyn Nasyedkin,[1,4,*] Isaac King,[2,*] Liangji Zhang,[1,*] Pei Chen,[2] Lili Wang,[2] Richard J. Staples,[3] Richard R. Lunt[1,2,**] and Johannes Pollanen[1,5‡]

[1]Department of Physics and Astronomy, Michigan State University, East Lansing, MI 48824, USA

[2]Department of Chemical Engineering and Materials Science, East Lansing, MI 48824, USA

[3]Department of Chemistry, Michigan State University, East Lansing, Michigan 48824, USA

[4]Neutron Scattering Division, Oak Ridge National Laboratory, Oak Ridge, TN 37831, USA

[5]Lead Contact

*These authors contributed equally to this work

**Correspondence: rlunt@msu.edu

‡Correspondence: pollanen@msu.edu

## SUMMARY

Inorganic halide perovskites have emerged as a promising platform in a wide range of applications from solar energy harvesting to computing, and light emission. The recent advent of epitaxial thin film growth of halide perovskites has made it possible to investigate low-dimensional quantum electronic devices based on this class of materials. This study leverages advances in vapor-phase epitaxy of halide perovskites to perform low-temperature magnetotransport measurements on single-domain cesium tin iodide ($CsSnI_3$) epitaxial thin films. The low-field magnetoresistance carries signatures of coherent quantum interference effects and spin-orbit coupling. These weak anti-localization measurements reveal a micron-scale low-temperature phase coherence length for charge carriers in this system. The results indicate that epitaxial halide perovskite heterostructures are a promising platform for investigating long coherent quantum electronic effects and potential applications in spintronics and spin-orbitronics.

## INTRODUCTION

Halide perovskite semiconductors are being developed worldwide in numerous thin-film electronic applications due to their exceptional properties. Initially these materials were investigated for light applications (Hattori et al., 1996) and subsequently for their photovoltaic properties (M. M. Lee et al., 2012; Nie et al., 2015; Yang et al., 2017), long carrier diffusion lengths (Saidaminov et al., 2020) and exceptional emission properties (Gao et al., 2020; Xu et al., 2019). More recently thin films of $CsSnX_3$ (X = Cl, Br, I) have also been used for lasers (Wang et al., 2015) and light-emitting diodes (LEDs) (Kazim et al., 2014; Yuan et al., 2016). Given their high-quality and flexible crystal structure it is natural to ask if halide perovskites might serve as a novel platform for creating quantum two-dimensional electron systems (2DESs). However, despite being considered as "defect-tolerant" films when grown from solution, there is still a need to transition these materials from polycrystalline layers to epitaxial single-crystal layers to enable the study of delicate and emergent quantum many-electron phenomena. By reducing the defect density in the perovskite layer, controlling strain and crystal phase, and improving crystalline order, electronic properties can be enhanced and quantum devices can be developed. Indeed, the ability to exploit epitaxy and epitaxial strain in silicon (Si) and III-V semiconductors revolutionized modern computing and optoelectronics and led to improved photovoltaics (Bertness et al., 1994; Dai et al., 2015; King et al., 2007; K. Lee et al., 2012; Takamoto et al., 1997), LEDs (Ghosh et al., 1986; Huang et al., 1992; Kuo et al., 2010; Peng and Wu, 2004) as well as 2DESs with ultra-high carrier mobility (Dingle et al., 1978; Hatke et al., 2017; Manfra, 2014; Pfeiffer et al., 1989; Umansky et al., 2009). For example, strained Si is now commonly found in every MOSFET (Sun et al., 2007) due to the increased interatomic distance in the silicon layer that nearly doubles the mobility. Additionally, molecular beam epitaxy (MBE) can now produce GaAs/AlGaAs heterostructures having charge carrier mobilities $\mu > 35 \times 10^6 cm^2/V \cdot s$ (Umansky et al., 2009) hosting a wide variety of exotic collective phases of quantum electronic matter (Eisenstein and Stormer, 1990). Advances in epitaxial growth have also benefitted a wide range of materials beyond low-bandgap semiconductors including graphene, diamond, SiC, and oxide perovskites just to name several.

The ability to precisely grow epitaxial halide perovskite thin films as well as halide perovskite quantum wells has recently been demonstrated (Chen et al., 2019, 2017; Oksenberg et al., 2020; L. Wang et al., 2017; Wang et al., 2019; Y. Wang et al., 2017). The vapor phase epitaxy of our $CsSnI_3$ samples was achieved on lattice-matched metal halide crystals with congruent ionic bonding (see Figure 1(a)) (Hu et al., 2017; L. Wang et al., 2017; Wang et al., 2019). The heteroepitaxy of these perovskite films allows for precise film thicknesses, controllable phase and orientation, integration into quantum wells, and opens the door to quantum transport devices. In fact, perovskite films have been shown to exhibit numerous quantum phenomena at low temperatures, such as superconductivity (Reyren et al., 2007), ferroelectricity (Cohen, 1992), and quantum coherent transport (Caviglia et al., 2010; Herranz et al., 2004; Keshavarz et al., 2017; Wang et al., 2020; Zhang et al., 2021), the last of which we show in this paper for epitaxial $CsSnI_3$ thin films. In particular, we exploit the growth of epitaxial halide perovskite devices to perform low-temperature quantum transport measurements on thin films composed of $CsSnI_3$. These measurements reveal quantum coherent transport of charge carriers in this halide perovskite 2DES as well as the presence of a spin-orbit crystal field.

In low-dimensional electronic systems quantum interference of charge carriers leads to a variety of coherent phenomena, including Aharonov-Bohm oscillations (Washburn and Webb, 1986), universal conductance fluctuations (Lee et al., 1987), as well quantum interference induced localization of charges (Altshuler et al., 1980; Bergmann, 1982; den dries et al., 1981; Pierre et al., 2003). Of particular interest are the effects of weak localization (WL) (Altshuler et al., 1980) and weak anti-localization (WAL) (Bergmann, 1982) as they can be used as tools to determine the phase coherence length for charge carriers in low-dimensional devices by performing magnetoresistance experiments at low temperature. Given a sufficiently long phase coherence length, the diffusive trajectory of a charge carrier in a low-dimensional system can interfere with its time-reversed partner. When the interference is <u>constructive</u> it produces weak localization, which manifests as a reduction in the electrical conductivity near zero magnetic field. In contrast, the presence of sufficiently strong spin-orbit coupling produces an additional rotation of the charge carrier spin and leads <u>destructive</u> interference, i.e. weak anti-localization. In this work we have observed clear signatures of phase coherent transport and WAL in the low-field magnetotransport measurements of epitaxial $CsSnI_3$. These results allow us to extract a phase coherence length of charge carriers in the device, which significantly exceeds the thin film thickness and indicates the low-dimensional character of the system.

## RESULTS AND DISCUSSION

The epitaxial $CsSnI_3$ crystalline film samples were grown in a custom Angstrom Engineering thermal evaporator by co-evaporation from two tungsten boats containing precursor materials (CsI and $SnI_2$), with each source having an independently calibrated quartz-crystal-microbalance rate monitor and source shutter. A 50 nm thick film was deposited stoichiometrically at a rate of 0.007 nm/s on a cleaved [100] surface of a potassium chloride (KCl) single crystal substrate. The growth was performed at pressures less than $3 \times 10^{-6}$ torr and a temperature of $22°C$. In-situ crystal analysis was performed in real-time using reflective high-energy electron diffraction (RHEED) to determine structure and film quality. The RHEED scans were performed at 30.0 kV and an emission current of less than 50 nA to reduce damage and charging on the perovskite film during growth. With these measurement conditions no damage was observed over typical deposition times of up to 1-2 hours. Gold contacts (50 nm thick) were deposited on the perovskite layers using electron-beam evaporation at a rate of 0.02 nm/s through a shadow mask in the same deposition chamber (with the same temperature and pressure).

The resulting $CsSnI_3$ epitaxial film is pseudomorphic to the KCl substrate ($a = 0.629\ nm$) as shown in Figure 1(a) and confirmed by rotation dependent RHEED patterns (Figure 1(b)) as well as x-ray pole figure analysis and TEM (see Supplemental Information section SI 1, Figure S1, S2 and S3). The streaky patterns observed are indicative of smooth growth compared to spotty patterns seen in rougher films (see Supplemental Information Figure S4 and S5). From the cleaved metal halide crystals, we observe atomic terrace lengths of 1-5 microns. Since the length of the streaks for the pseudomorphic epitaxial layer is similar to the streak lengths from the substrate, this indicates that the terrace length is simply limited by that of substrate. The patterns we observe are representative of high quality epitaxial and heteroepitaxial oxide perovskite layers(Choi et al., 2012; Soukiassian et al., 2008). We note that it is possible that a small amount of oxygen and water impurities could be incorporated into the epitaxial thin film despite the overall low pressure during growth. In fact, future investigations systematically varying the growth pressure could be conducted to assess the impact of the base pressure on the resulting transport properties of the epitaxial thin film.

The RHEED data show a clear change in the symmetry from face-centered-cubic (FCC) of the substrate to a primitive perovskite cell by the emergence of (01) streaks between the (02) streaks of the substrate. The crystal structure of epitaxial $CsSnI_3$ is similar to the high temperature cubic phase (L. Wang et al., 2017) but tetragonally distorted due to the in-plane tensile strain resulting in $c = 0.612 \pm 0.002\ nm$ (see Supplemental Information, Figure S6 and Table S1). Thus, this epitaxial phase and band structure are unique from the commonly observed orthorhombic phase at room temperature (L. Wang et al., 2017).

Two $2.0\ mm \times 5.0\ mm$ gold pads spaced by $50\ \mu m$ on the $CsSnI_3$ layer enabled standard low-frequency ($10\ Hz$) ac electrical transport measurements as shown in Figure 1(c). While the epitaxial perovskite film is more stable than the bulk orthorhombic phase in air, it too degrades when exposed to air after many hours, so protocols were developed to keep the devices in a high vacuum or dry-nitrogen filled environment, similar to a glovebox, during sample fabrication, transferring, and measurement. Additionally we have characterized the air sensitivity of epitaxial $CsSnI_3$ using x-ray, optical and electrical transport techniques (see Supplemental Information section SI 2, Figure S7). We note that it is conceivable that for particular future applications the material could be encapsulated in epoxy or edge sealed in an inert atmosphere after growth to prevent degradation in air. Alternatively the material could be kept under vacuum, which is the typical environment for cryogenic quantum devices at mK temperatures. Devices for the quantum transport measurements reported here were mounted in a custom indium o-ring sealed copper sample holder (see Supplemental Information Figure S8), to minimize exposure to air, and thermally anchored to the mixing chamber of a dilution refrigerator having a lowest temperature of $T{\sim}10\ mK$. A variable magnetic field $B$ perpendicular to the plane of the CsSnI$_3$ epitaxial film was supplied by a superconducting solenoid and enabled the measurement of the magnetoconductivity of the device up to $B = \pm 13.5\ T$.

Before investigating the magnetotransport, we first characterized the temperature dependent electrical properties of the device in the absence of a magnetic field (for comparison we present measurements on additional CsSnI$_3$ devices which show consistent behavior as discussed below. Similar measurements on epitaxial CsSnBr$_3$ devices were also performed. See Supplemental Information section SI 3, Figure S9). In Figure 2 we show the conductivity, $\sigma$, of a CsSnI$_3$ epitaxial film as the sample was cooled from room temperature down to $T = 16\ mK$. The resistance was measured using standard lock-in techniques and the conductivity was calculated from the measured device resistance by taking into account the geometric factor between the gold pads. The low overall measured resistance indicates that the Au makes ohmic contact to the CsSnI$_3$, consistent with previous electrical measurements on bulk CsSnI$_3$ (Han et al., 2019). Since the total measured resistance is given by $R_{tot} = 2R_c + R$, where $R$ is the intrinsic sample resistance, these measurements place an upper bound of $R_c \approx 10\ \Omega$ on the contact resistance between the gold pads and the thin film and indicate good electrical contact between the two. The conductivity of the device increases with decreasing temperature indicating the presence of mobile charge carriers that do not freeze out upon cooling to low temperature. This increase in the conductivity is consistent with a reduction in phonon density upon lowering the temperature of the device. The existence of charge carriers in the material is likely the result of intrinsic doping in the epitaxial thin film, which is a direct bandgap semiconductor (see Supplemental Information section SI 4, Figure S10 and Table S2). In fact, previous electrical conductivity measurements on orthorhombic CsSnI$_3$ have also reported metal-like conduction attributable to hole-doping associated with Sn vacancies (Chung et al., 2012). We note that it is possible the magnitude of the low temperature conductivity shown in Figure 2 is limited by the contact resistance at the Au/CsSnI$_3$ interface even though there is clearly good ohmic contact. This would impact the shape of the temperature dependence of $\sigma$ as well as its maximum value but would not change the conclusions we draw regarding phase coherent transport nor the existence of spin-orbit coupling described below.

At low-temperatures, the application of a magnetic field normal to the plane of the CsSnI$_3$ layer (see Figure 1(c)) strongly modifies the transport of charge carriers. In Figure 3(a) we present the conductivity of the CsSnI$_3$ epitaxial film device as a function of magnetic field up to $\pm 13.5\ T$. Above several Tesla we observe a non-saturating magnetoconductivity that is approximately linear in the applied field. This apparent reduction in the conductivity with increasing magnetic field likely arises from a relatively mundane source, i.e. from a convolution of the longitudinal conductivity measurement by the Hall voltage across the sample due to the geometry of our device. In fact, this type of effect is common in magnetotransport measurements such as ours and has been observed in 2DESs in semiconductors(Fang and Stiles, 1983; Russell et al., 1990) as well as graphene (Skachko et al., 2010), and is associated with the formation of so-called "hot-spots" near corners of device contacts (Klaß et al., 1991; Russell et al., 1990; Skachko et al., 2010).

More importantly, the low-field magnetoconductivity can be used to reveal signatures of coherent quantum phenomena. In the vicinity of $B = 0$ we observe clear signs of quantum interference. The inset of Figure 3(a) shows the change in the magnetoconductance $\sigma(B) - \sigma(0)$ near zero magnetic field and $T = 16\ mK$. These data exhibit a cusp-like maximum that broadens and weakens as the temperature is increased as shown in Figure 3(b). As we describe below, these features are the characteristic signatures of weak anti-localization, i.e. phase coherent diffusive magnetotransport in the presence of spin-orbit coupling (Bergmann, 1982; Lee and Ramakrishnan, 1985).

The Hikami-Larkin-Nagaoka (HLN) theory of weak anti-localization provides a description of the magnetoconductivity behavior of quasi-2D systems in the quantum diffusive transport regime (Hikami et al., 1980). We utilize a simplified empirical HLN formula to analyze our data and to extract a phase coherence length $L_\Phi$ for the charge carriers. This model contains two fit parameters $L_\Phi$ and $\alpha$, which accurately capture the physics of WAL in our data. In this analysis, the change in magnetoconductivity away from its zero field value is given by,

$$\sigma(B) - \sigma(0) = \frac{\alpha}{\pi}\left[\Psi\left(\frac{1}{2} + \frac{\hbar}{4eBL_\Phi^2}\right) - \ln\left(\frac{\hbar}{4eBL_\Phi^2}\right)\right] \quad (1)$$

where $e$ is the electron charge, $\hbar$ is the reduced Planck's constant and $\Psi(x)$ is the digamma function. The sign of the coefficient $\alpha$ indicates the type of localization (He et al., 2011; Kurzman et al., 2011). Specifically, $\alpha > 0$ is associated with WL while $\alpha < 0$ indicates WAL. For the entire range over which we observe phase coherent transport ($T < 15K$) we find a negative value of $\alpha$, which demonstrates the presence of spin-orbit coupling in the epitaxial CsSnI$_3$ film. Representative fits of the data to Equation 1 at different temperatures are shown in Figure 3(b). The magnitude of spin-orbit coupling in a material is dictated by the mass of the constituent atoms in the crystal and whether the crystal lacks inversion symmetry. Inversion symmetry is broken if a compound is intrinsically non-centrosymmetric and results in Dresselhaus spin-orbit coupling (Dresselhaus, 1955). It is not known whether the epitaxial phase of CsSnI$_3$ intrinsically lacks an inversion center as it is nearly impossible to perform precise atomic refinement on an epitaxial thin film. Regardless, the presence of the interface between the KCl substrate and the CsSnI$_3$ epitaxial film breaks inversion symmetry and could lead to a Rashba spin-orbit field (Bychkov and Rasbha, 1984) along the growth direction of the heterostructure. In fact, it has been shown that the large Rashba splitting observed in tetragonal MAPbI$_3$ is likely not a bulk-, but rather a surface-induced phenomenon (Frohna et al., 2018). Alternatively, it is possible that the spin-orbit coupling in CsSnI$_3$ that we observe arises from the relatively large mass of the constituent atoms in the crystal. We note that our quasi-two-terminal measurements do not provide a measure of the absolute sample conductivity, which could be affected by the contact resistance. Therefore in our analysis the magnitude of $\alpha$ serves only as a overall scale factor needed to accurately fit the data and to extract the phase coherence length. However, as described above, the sign of $\alpha$ does carry intrinsic physical information, i.e. that spin-orbit coupling is present in the system. Additionally we emphasize that in this analysis $\alpha$ and $L_\Phi$ are not meaningfully dependent on one another. This can be seen in Equation 1 where $\alpha$ is an overall scaling pre-factor that cannot simply be subsumed into the digamma and log functions to compensate for a change in $L_\Phi$.

The temperature dependence of the phase coherence length extracted from these measurements is shown in Figure 4 along with value of $L_\Phi$ for other low-dimensional quantum electronic materials in the quantum diffusive transport regime. For our data on epitaxial CsSnI$_3$ each value of $L_\Phi$ corresponds to the average value obtained from multiple measurements at each temperature and the error bars are the standard deviation. As the temperature is increased we observe a reduction in $L_\Phi$ due to increased inelastic scattering of charge carriers. Several important inelastic scattering mechanisms can result in the loss of phase coherence including electron-electron scattering, electron-phonon scattering, and scattering from magnetic impurities (Lin and Bird, 2002). In the quantum diffusive regime $L_\Phi = \sqrt{D\tau_\Phi}$, where $D$ is the charge carrier diffusivity and $\tau_\Phi$ is the rate of dephasing scattering events. The temperature dependence of $\tau_\Phi$ carries information regarding which of the scattering mechanisms dominates, however multiple competing mechanisms often lead to mixed temperature dependences that are difficult to disentangle (Lin and Bird, 2002). We find that $\tau_\Phi$ scales as $1/T^p$ with $p \simeq 2$, which is consistent with phonon scattering as the dominant contributor to dephasing with increasing temperature. In fact, previous studies on semiconductors and metals have reported electron-phonon scattering as leading to $p \simeq 2 - 4$ (Lin and Bird, 2002). Regarding the role of magnetic scattering, while we do not purposefully introduce magnetic defects/impurities (e.g. Fe, Mn, etc.) magnetic scattering is not impossible in our devices. While the purity of the source materials is high (CsI 99.9% (Sigma-Aldrich), SnI$_2$ 99+% (Alfa Aesar), SnI$_2$ 99% (Strem)) it is possible for a residual amount of magnetic impurities to be incorporated during growth. Additionally, it has recently been demonstrated that vacancy-induced magnetism can arise in solution processed lead halide perovskites (Sun et al., 2020). Future studies could be conducted on other epitaxial halide perovskites, such as CsSnI$_3$, to systematically investigate the role of magnetism on quantum transport devices based on these materials. Regardless of the underlying inelastic scattering mechanism $L_\Phi$ increases with decreasing temperature and reaches $\approx 5\ \mu m$ below roughly $350\ mK$, below which it saturates, possibly due to thermal decoupling of the charge carriers at low temperatures. We note that recent work on three-dimensional quasi-epitaxial layers of CsPbBr$_3$ have reported weak localization signatures in photoelectric transport with $L_\Phi \approx 50\ nm$ at $3.5\ K$ (Wang et al., 2020), likely limited by the quasi-epitaxial domain size.

The value of the charge carrier phase coherence length we find in epitaxial CsSnI$_3$ can be compared to other high-quality low-dimensional electronic materials in the quantum diffusive regime where localization measurements are possible such as GaAs (Beenakker and van Houten, 1991; Mailly, 1987), oxide thin films (Yun et al., 2017), graphene (Baker et al., 2012; Elias and Henriksen, 2017; Ki et al., 2008; Tikhonenko et al., 2008) as well as silicon (Beenakker and van Houten, 1991). We note that longer phase coherence exists in extremely high-mobility GaAs 2DESs and graphene but are not reported due to the absence of localization effects in sufficiently disorder free samples. In fact, in graphene ballistic transport has been observed over longer than 15 microns (Wang et al., 2013).

In summary, our results on phase coherent transport in the presence of spin-orbit coupling show that micron-scale phase coherence lengths can be achieved in epitaxial CsSnI$_3$. These findings attest to the fact that epitaxial halide perovskites are emerging as a novel material for future quantum coherent devices with potential applications in spin-orbitronics where gate-control could enable manipulation of spin via the Rashba-effect (Kepenekian and Even, 2017; Manchon et al., 2015).

### LIMITATIONS OF THE STUDY

As described in the main text of the manuscript, transport measurements were performed using a two-terminal configuration. These measurements reveal, via the manifestation of weak anti-localization, spin-orbit coupling and long phase coherent transport of charge carriers. Future work using a four-terminal van der Pauw geometry would allow for measurements of the zero-field mobility needed to characterize the elastic mean-free path of charge carriers in this material and four-terminal measurements of the Hall resistance as a function of magnetic field would allow characterization of the type and density of charge carriers. These measurements could be performed to extensively map out the dependence of the mobility and density on various epitaxial growth parameters, doping level, disorder, etc. Such future measurements would be ideal for optimizing sample growth with an eye toward quantum electronic device applications as they could be used to understand how to develop samples with decreasing levels of impurity scattering. Additionally such measurements would allow for a quantitative measure of the strength of the spin-orbit coupling revealed by our two-terminal measurements.

## ACKNOWLEDGMENTS

We are grateful to E.A. Henriksen, D.G. Schlom, N.O. Birge, M.I. Dykman, S.D. Mahanti, J.I.A. Li, A. Sen and J.R. Lane for illuminating and fruitful discussions. We also thank R. Loloee and B. Bi for technical assistance and use of the W. M. Keck Microfabrication Facility at MSU. This work was supported by the National Science Foundation via grant no. DMR-1807573. J. Pollanen also acknowledges the valuable support of the Cowen Family Endowment at MSU. This research used resources of the Advanced Photon Source, a U.S. Department of Energy (DOE) Office of Science User Facility operated for the DOE Office of Science by Argonne National Laboratory under Contract No. DE-AC02-06CH11357. Use of the LS-CAT Sector 21 was supported by the Michigan Economic Development Corporation and the Michigan Technology Tri-Corridor (Grant 085P1000817). Data was collected at the Life Sciences Collaborative Access Team beamline 21-ID-D at the Advanced Photon Source, Argonne National Laboratory. We acknowledge Dr. Zdzislaw Wawrzak for his help making the temperature dependent synchrotron measurements.

## AUTHOR CONTRIBUTIONS

K.N. and L.Z. performed the low-temperature magneto-transport experiments and I.K. and P.C. grew the halide perovskite films and devices. I.K., P.C., L.W., and R.J.S. performed x-ray characterization of the samples. J.P. and R.R.L. conceived of the experiments and supervised the project. All authors contributed to data analysis and writing the manuscript.

## DECLARATION OF INTERESTS

The authors declare no competing interests.

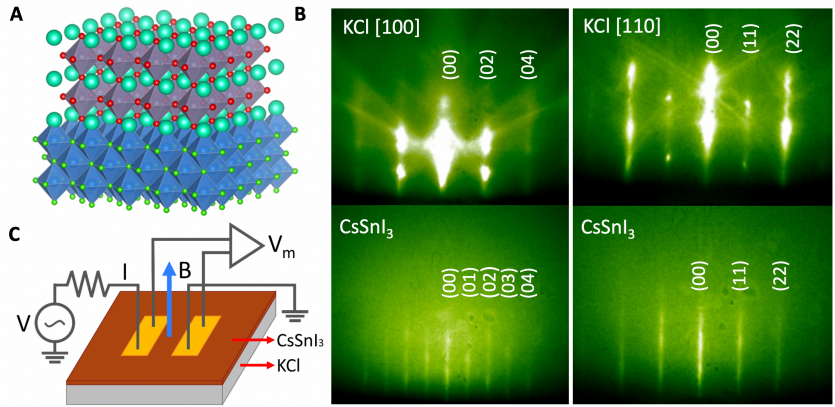

Figure 1. Crystal structure, RHEED pattern and schematic of the experimental setup

(A) Crystal structure of epitaxial cesium tin iodide (CsSnI$_3$) on potassium chloride (KCl) (Cyan is Cs, Gray is Sn, Red is I, Green is K, Blue is Cl). (B) Reflective high-energy electron diffraction (RHEED) pattern of the KCl substrate (top panels) and CsSnI$_3$ thin film showing well-defined crystalline streaks in both the KCl and epitaxially grown CsSnI$_3$. The appearance of (01) and (03) streaks indicate the transformation from the FCC lattice of KCl to the primitive lattice of the perovskite.
(C) Schematic of the experimental setup. Electrical transport measurements were performed between evaporated gold contacts on the 50~nm thick epitaxial CsSnI$_3$ thin film devices grown on a cleaved KCl substrate. The conductivity was calculated from the measured value of the voltage $V_m$ and the sourced current $I$ taking into account the geometric factor between the gold contacts. A magnetic field, $B$, perpendicular to the plane of the CsSnI$_3$ enabled measurement of the magnetotransport at low temperature.

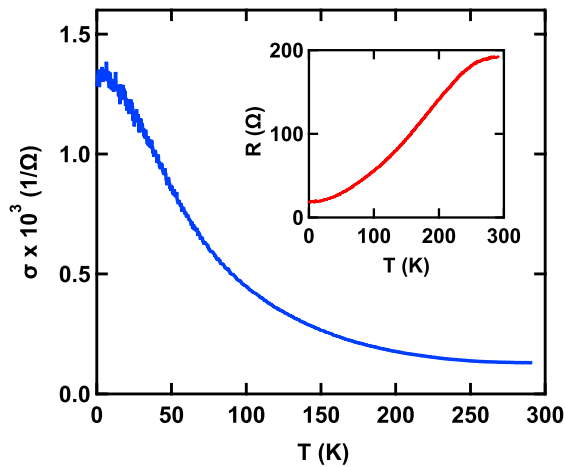

Figure 2. Electrical conductivity of epitaxial CsSnI$_3$

Electrical conductivity $\sigma$ of a epitaxial CsSnI$_3$ thin film as a function of temperature with $B = 0$. Inset: the corresponding device resistance $R$ versus temperature.

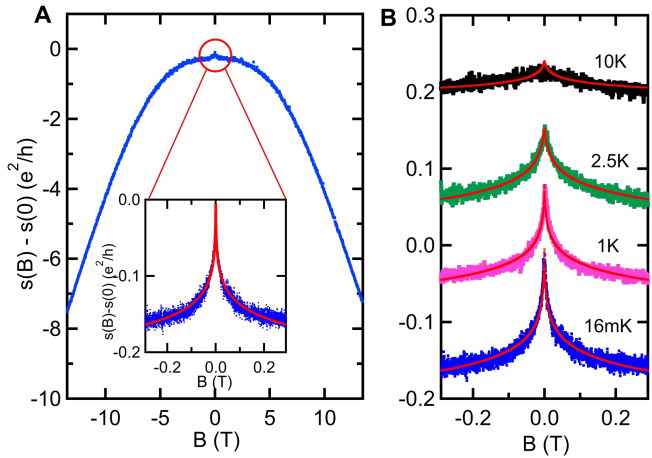

### Figure 3. Magnetoconductivity of epitaxial CsSnI$_3$

(**A**) Magnetoconductivity of the epitaxial CsSnI$_3$ thin film as a function of magnetic field normal to the plane of the device at $T = 16\ mK$. The vertical axis is the magnetoconductivity $\sigma(B)$ minus the value of the conductivity at zero magnetic field $\sigma(0)$. Inset: low field magnetoconductivity showing a clear weak anti-localization peak at $B = 0$. The solid line is a fit to the WAL theory, Equation 1, described in the text, from which we can extract the charge carrier coherence length $L_\phi$.

(**B**) As the temperature is increased the WAL peak broadens and weakens as $L_\phi$ decreases due to increased inelastic scattering of charge carriers.

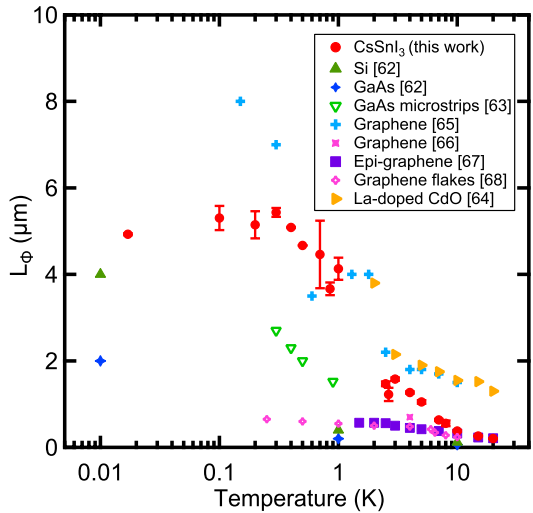

### Figure 4. Phase coherence length

Temperature dependence of the charge carrier phase coherence length $L_\Phi$ in the epitaxial CsSnI$_3$ thin film. Each value of $L_\Phi$ corresponds to the average value obtained from multiple measurements at a given temperature and the error bars are the standard deviation. For comparison we also show results from the literature for other electronic materials.

# STAR METHODS

## KEY RESOURCES TABLE

| REAGENT or RESOURCE | SOURCE | IDENTIFIER |
|---|---|---|
| Software | | |
| Mathematica | https://www.wolfram.com/mathematica/ | version 12.3 |
| Materials Studio | https://www.3ds.com/products-services/biovia/products/molecular-modeling-simulation/biovia-materials-studio/ | version 7.0 |
| | | |

## RESOURCE AVAILABILITY

### Lead contact

Further information and request for resources should be directed to the lead contact, Johannes Pollanen (pollanen@msu.edu)

### Materials availability

This study did not generate new unique reagents.

### Data and code availability

The data are available upon reasonable request by contacting the lead contact. No new code was generated during the course of this study.

## METHOD DETAILS

### $CsSnI_3$ sample growth

The epitaxial $CsSnI_3$ crystalline film samples were grown in a custom Angstrom Engineering thermal evaporator by co-evaporation from two tungsten boats containing precursor materials (CsI and $SnI_2$), with each source having an independently calibrated quartz-crystal-microbalance rate monitor and source shutter. A 50 nm thick film was deposited stoichiometrically at a rate of 0.007 nm/s on a cleaved [100] surface of a potassium chloride (KCl) single crystal substrate. The growth was performed at pressures less than $3 \times 10^{-6}$ torr and a temperature of $22°C$. In-situ crystal analysis was performed in real-time using reflective high-energy electron diffraction (RHEED) to determine structure and film quality. The RHEED scans were performed at 30.0 kV and an emission current of less than 50 nA to reduce damage and charging on the perovskite film during growth. With these measurement conditions no damage was observed over typical deposition times of up to 1-2 hours. Gold contacts (50 nm thick) were deposited on the perovskite layers using electron-beam evaporation at a rate of 0.02 nm/s through a shadow mask in the same deposition chamber (with the same temperature and pressure).

### Characterization of $CsSnI_3$ air sensitivity and device handling

To understand the effects of exposure to air, we have characterized the air sensitivity of additional epitaxial $CsSnI_3$ devices using x-ray, optical and electrical transport, as shown in Figure. S7. In Figure. S7(c) we show how the resistance of a $CsSnI_3$ epitaxial layer evolves as a function of time when the device was exposed in air at room temperature. For the first 15-20 minutes, the device resistance does not significantly change. With continued exposure we clearly observe an increase in the device resistance with increasing time, which continues until it is > 50 MΩ. We attribute this behavior to the known transformation of $CsSnI_3$ into $Cs_2SnI_6$ upon exposure to air (even for the epitaxial phase). Importantly for the results reported in this manuscript no such changes in device resistance were observed for any of the devices that were handled and sealed in a dry-nitrogen environment before being measured at low temperature. Moreover, repeated transport measurements performed at low temperature did not show changes in the overall resistance of the device nor changes in the observed magnetotransport phenomena.

Although the epitaxial films were structurally stable in air for hours, care was taken to mitigate this as a convoluting effect. Thus the growth, handling and characterization of the devices in this article ($CsSnI_3$ and $CsSnBr_3$) were performed, from start to finish, in a dry, oxygen-free environment. After the fabrication, the devices were sealed in containers filled with dry nitrogen gas and transported from the growth laboratory to the low temperature quantum measurement laboratory. These devices were wired up in a nitrogen environment and loaded into a custom hermetically sealed sample cell containing an 18-pin chip carrier inside (see Figure. S8). The cell was sealed with a conventional indium O-ring compatible with cryogenic measurements. Then the cell was transferred to dilution refrigerator for low temperature (~10 mK) and high magnetic field (~14 T) transport measurements. At no point during this entire process (from growth to low-temperature measurement) were the devices exposed to air.

*Low-temperature magneto-transport measurements*

After growth the devices were wired up using 0.02″ diameter gold wire and InSn solder in a nitrogen environment and loaded into a custom hermetically sealed sample cell containing an 18-pin chip carrier inside This experimental cell was sealed with a conventional indium O-ring compatible with cryogenic measurements. Then the cell was transferred to cryogen-free dilution refrigerator for low temperature (~10 mK) and high magnetic field (~14 T) transport measurements which were performed using low-frequency (10 Hz) lock-in techniques. At no point during this entire process (from growth to low-temperature measurement) were the devices exposed to ambient condition.

## QUANTIFICATION AND STATISTICAL ANALYSIS

Each value of $L_\Phi$ shown in Figure. 4 of the main manuscript corresponds to the average value obtained from multiple weak-anti-localization measurements at a given temperature and the error bars are the standard deviation.

## REFERENCES

Altshuler, B.L., Khmel'nitzkii, D., Larkin, A.I., Lee, P.A. (1980). Magnetoresistance and Hall effect in a disordered two-dimensional electron gas. Phys. Rev. B 22, 5142–5153. https://doi.org/10.1103/PhysRevB.22.5142

Baker, A.M.R., Alexander-Webber, J.A., Altebaeumer, T., Janssen, T.J.B.M., Tzalenchuk, A., Lara-Avila, S., Kubatkin, S., Yakimova, R., Lin, C.-T., Li, L.-J., et al. (2012). Weak localization scattering lengths in epitaxial, and CVD graphene. Phys. Rev. B 86, 235441. https://doi.org/10.1103/PhysRevB.86.235441

Beenakker, C.W.J., van Houten, H. (1991). Quantum Transport in Semiconductor Nanostructures, in: Ehrenreich, H., Turnbull, D. (Eds.), Semiconductor Heterostructures and Nanostructures, Solid State Physics. Academic Press, pp. 1–228. https://doi.org/10.1016/S0081-1947(08)60091-0

Bergmann, G. (1982). Weak anti-localization---An experimental proof for the destructive interference of rotated spin 12. Solid State Commun. 42, 815–817. https://doi.org/10.1016/0038-1098(82)90013-8

Bertness, K.A., Kurtz, S.R., Friedman, D.J., Kibbler, A.E., Kramer, C., Olson, J.M. (1994). 29.5%-efficient GaInP/GaAs tandem solar cells. Appl. Phys. Lett. 65, 989–991. https://doi.org/10.1063/1.112171

Bychkov, Y.A., Rasbha, E.I. (1984). Properties of a 2D electron gas with lifted spectral degeneracy. P. Zh. Eksp. Teor. Fiz. 39, 66–69

Caviglia, A.D., Gabay, M., Gariglio, S., Reyren, N., Cancellieri, C., Triscone, J.-M. (2010). Tunable Rashba spin-orbit interaction at oxide interfaces. Phys. Rev. Lett. 104, 126803. https://doi.org/10.1103/PhysRevLett.104.126803

Chen, J., Luo, Z., Fu, Y., Wang, X., Czech, K.J., Shen, S., Guo, L., Wright, J.C., Pan, A., Jin, S. (2019). Tin (IV)-tolerant vapor-phase growth and photophysical properties of aligned cesium tin halide perovskite ($CsSnX_3$; X= Br, I) nanowires. ACS Energy Lett. 4, 1045–1052. https://doi.org/10.1021/acsenergylett.9b00543

Chen, J., Morrow, D.J., Fu, Y., Zheng, W., Zhao, Y., Dang, L., Stolt, M.J., Kohler, D.D., Wang, X., Czech, K.J., et al. (2017). Single-crystal thin films of cesium lead bromide perovskite epitaxially grown on metal oxide perovskite ($SrTiO_3$). J. Am. Chem. Soc. 139, 13525–13532. https://doi.org/10.1021/jacs.7b07506

Choi, M., Posadas, A., Dargis, R., Shih, C.-K., Demkov, A.A., Triyoso, D.H., David Theodore, N., Dubourdieu, C., Bruley, J., Jordan-Sweet, J. (2012). Strain relaxation in single crystal $SrTiO_3$ grown on Si (001) by molecular beam epitaxy. J. Appl. Phys. 111, 64112. https://doi.org/10.1063/1.3695998

Chung, I., Song, J.-H., Im, J., Androulakis, J., Malliakas, C.D., Li, H., Freeman, A.J., Kenney, J.T., Kanatzidis, M.G. (2012). $CsSnI_3$: Semiconductor or Metal? High Electrical Conductivity and Strong Near-Infrared Photoluminescence from a Single Material. High Hole Mobility and Phase-Transitions. J. Am. Chem. Soc. 134, 8579–8587. https://doi.org/10.1021/ja301539s

Cohen, R.E. (1992). Origin of ferroelectricity in perovskite oxides. Nature 358, 136–138. https://doi.org/10.1038/358136a0

Dai, P., Tan, M., Wu, Y.Y., Ji, L., Bian, L.F., Lu, S.L., Yang, H. (2015). Solid-state tellurium doping of AlInP and its application to photovoltaic devices grown by molecular beam epitaxy. J. Cryst. Growth 413, 71–75. https://doi.org/10.1016/j.jcrysgro.2014.12.014

den dries, L., Van Haesendonck, C., Bruynseraede, Y., Deutscher, G. (1981). Two-Dimensional Localization in Thin Copper Films. Phys. Rev. Lett. 46, 565–568. https://doi.org/10.1103/PhysRevLett.46.565

Dingle, R., Störmer, H.L., Gossard, A.C., Wiegmann, W. (1978). Electron mobilities in modulation-doped semiconductor heterojunction superlattices. Appl. Phys. Lett. 33, 665–667. https://doi.org/10.1063/1.90457

Dresselhaus, G. (1955). Spin-Orbit Coupling Effects in Zinc Blende Structures. Phys. Rev. 100, 580–586. https://doi.org/10.1103/PhysRev.100.580

Eisenstein, J.P., Stormer, H.L. (1990). The fractional quantum Hall effect. Science (80-. ). 248, 1510–1516. https://doi.org/10.1126/science.248.4962.1510

Elias, J.A., Henriksen, E.A. (2017). Electronic transport and scattering times in tungsten-decorated graphene. Phys. Rev. B 95, 75405. https://doi.org/10.1103/PhysRevB.95.075405

Fang, F.F., Stiles, P.J. (1983). Quantized magnetoresistance in two-dimensional electron systems. Phys. Rev. B 27, 6487–6488. https://doi.org/10.1103/PhysRevB.27.6487

Frohna, K., Deshpande, T., Harter, J., Peng, W., Barker, B.A., Neaton, J.B., Louie, S.G., Bakr, O.M., Hsieh, D., Bernardi, M. (2018). Inversion symmetry and bulk Rashba effect in methylammonium lead iodide perovskite single crystals. Nat. Commun. 9, 1829. https://doi.org/10.1038/s41467-018-04212-w

Gao, L., Quan, L.N., de Arquer, F.P., Zhao, Y., Munir, R., Proppe, A., Quintero-Bermudez, R., Zou, C., Yang, Z., Saidaminov, M.I., et al. (2020). Efficient near-infrared light-emitting diodes based on quantum dots in layered perovskite. Nat. Photonics. https://doi.org/10.1038/s41566-019-0577-1

Ghosh, R.N., Griffing, B., Ballantyne, J.M. (1986). Monolithic integration of GaAs light-emitting diodes and Si metal-oxide-semiconductor field-effect transistors. Appl. Phys. Lett. 48, 370–371. https://doi.org/10.1063/1.96555

Groom, C.R., Bruno, I.J., Lightfoot, M.P., Ward, S.C. (2016). The Cambridge structural database. Acta Crystallogr. Sect. B Struct. Sci. Cryst. Eng. Mater. 72, 171–179. https://doi.org/10.1107/S2052520616003954

Han, J.S., Le, Q. Van, Choi, J., Kim, H., Kim, S.G., Hong, K., Moon, C.W., Kim, T.L., Kim, S.Y., Jang, H.W. (2019). Lead-Free All-Inorganic Cesium Tin Iodide Perovskite for Filamentary and Interface-Type Resistive Switching toward Environment-Friendly and Temperature-Tolerant Nonvolatile Memories. ACS Appl. Mater. Interfaces 11, 8155–8163. https://doi.org/10.1021/acsami.8b15769

Hatke, A.T., Wang, T., Thomas, C., Gardner, G.C., Manfra, M.J. (2017). Mobility in excess of $10^6$ $cm^2/V$ s in InAs quantum wells grown on lattice mismatched InP substrates. Appl. Phys. Lett. 111, 142106. https://doi.org/10.1063/1.4993784

Hattori, T., Taira, T., Era, M., Tsutsui, T., Saito, S. (1996). Highly efficient electroluminescence from a heterostructure device combined with emissive layered-perovskite and an electron-transporting organic compound. Chem. Phys. Lett. 254, 103–108. https://doi.org/10.1016/0009-2614(96)00310-7

He, H.-T., Wang, G., Zhang, T., Sou, I.-K., Wong, G.K.L., Wang, J.-N., Lu, H.-Z., Shen, S.-Q., Zhang, F.-C. (2011). Impurity Effect on Weak Antilocalization in the Topological Insulator $Bi_2Te_3$. Phys. Rev. Lett. 106, 166805. https://doi.org/10.1103/PhysRevLett.106.166805

Herranz, G., Sánchez, F., Martinez, B., Fontcuberta, J., Garcia-Cuenca, M. V, Ferrater, C., Varela, M., Levy, P. (2004). Weak localization effects in some metallic perovskites. Eur. Phys. J. B-Condensed Matter Complex Syst. 40, 439–444. https://doi.org/10.1140/epjb/e2004-00207-9

Hikami, S., Larkin, A.I., Nagaoka, Y. (1980). Spin-Orbit Interaction and Magnetoresistance in the Two Dimensional Random System. Prog. Theor. Phys. 63, 707–710. https://doi.org/10.1143/PTP.63.707

Hu, Y., Guo, Y., Wang, Y., Chen, Z., Sun, X., Feng, J., Lu, T.-M., Wertz, E., Shi, J. (2017). A review on low dimensional metal halides: Vapor phase epitaxy and physical properties. J. Mater. Res. 32, 3992–4024. https://doi.org/10.1557/jmr.2017.325

Huang, K.H., Yu, J.G., Kuo, C.P., Fletcher, R.M., Osentowski, T.D., Stinson, L.J., Craford, M.G., Liao, A.S.H. (1992). Twofold efficiency improvement in high performance AlGaInP light-emitting diodes in the 555-620 nm spectral region using a thick GaP window layer. Appl. Phys. Lett. 61, 1045–1047. https://doi.org/10.1063/1.107711

Kazim, S., Nazeeruddin, M.K., Grätzel, M., Ahmad, S. (2014). Perovskite as Light Harvester: A Game Changer in Photovoltaics. Angew. Chemie Int. Ed. 53, 2812–2824. https://doi.org/10.1002/anie.201308719

Kepenekian, M., Even, J. (2017). Rashba and Dresselhaus Couplings in Halide Perovskites: Accomplishments and Opportunities for Spintronics and Spin-Orbitronics. J. Phys. Chem. Lett. 8, 3362–3370. https://doi.org/10.1021/acs.jpclett.7b01015

Keshavarz, M., Wiedmann, S., Yuan, H., Debroye, E., Roeffaers, M., Hofkens, J. (2017). Light-and Temperature-Modulated Magneto-Transport in Organic-Inorganic Lead Halide Perovskites. ACS Energy Lett. 3, 39–45. https://doi.org/10.1021/acsenergylett.7b00941

Ki, D.-K., Jeong, D., Choi, J.-H., Lee, H.-J., Park, K.-S. (2008). Inelastic scattering in a monolayer graphene sheet: A weak-localization study. Phys. Rev. B 78, 125409. https://doi.org/10.1103/PhysRevB.78.125409

King, R.R., Law, D.C., Edmondson, K.M., Fetzer, C.M., Kinsey, G.S., Yoon, H., Sherif, R.A., Karam, N.H. (2007). 40% efficient metamorphic GaInP/GaInAs/Ge multijunction solar cells. Appl. Phys. Lett. 90, 183516. https://doi.org/10.1063/1.2734507

Klaß, U., Dietsche, W., von Klitzing, K., Ploog, K. (1991). Imaging of the dissipation in quantum-Hall-effect experiments. Zeitschrift für Phys. B Condens. Matter 82, 351–354. https://doi.org/10.1007/BF01357178

Kuo, D.-M., Wang, S.-J., Uang, K.-M., Chen, T.-M., Lee, W.-C., Wang, P.-R. (2010). Enhanced Light Output of AlGaInP Light Emitting Diodes Using an Indium-Zinc Oxide Transparent Conduction Layer and Electroplated Metal Substrate. Appl. Phys. Express 4, 12101. https://doi.org/10.1143/apex.4.012101

Kurzman, J.A., Miao, M.-S., Seshadri, R. (2011). Hybrid functional electronic structure of $PbPdO_2$, a small-gap semiconductor. J. Phys. Condens. Matter 23, 465501. https://doi.org/10.1088/0953-8984/23/46/465501

Lee, K., Zimmerman, J.D., Xiao, X., Sun, K., Forrest, S.R. (2012). Reuse of GaAs substrates for epitaxial lift-off by employing protection layers. J. Appl. Phys. 111, 33527. https://doi.org/10.1063/1.3684555

Lee, M.M., Teuscher, J., Miyasaka, T., Murakami, T.N., Snaith, H.J. (2012). Efficient Hybrid Solar Cells Based on Meso-Superstructured Organometal Halide Perovskites. Science (80-. ). 338, 643–647. https://doi.org/10.1126/science.1228604

Lee, P.A., Ramakrishnan, T. V (1985). Disordered electronic systems. Rev. Mod. Phys. 57, 287–337. https://doi.org/10.1103/RevModPhys.57.287

Lee, P.A., Stone, A.D., Fukuyama, H. (1987). Universal conductance fluctuations in metals: Effects of finite temperature, interactions, and magnetic field. Phys. Rev. B 35, 1039–1070. https://doi.org/10.1103/PhysRevB.35.1039

Lin, J.J., Bird, J.P. (2002). Recent experimental studies of electron dephasing in metal and semiconductor mesoscopic structures. J. Phys. Condens. Matter 14, R501--R596. https://doi.org/10.1088/0953-8984/14/18/201

Mailly, D. (1987). Weak Localization and Interaction Corrections in

Microstrips of GaAs. Europhys. Lett. 4, 1171–1176. https://doi.org/10.1209/0295-5075/4/10/015

Manchon, A., Koo, H.C., Nitta, J., Frolov, S.M., Duine, R.A. (2015). New perspectives for Rashba spin-orbit coupling. Nat. Mater. 14, 871–882. https://doi.org/10.1038/nmat4360

Manfra, M.J. (2014). Molecular Beam Epitaxy of Ultra-High-Quality AlGaAs/GaAs Heterostructures: Enabling Physics in Low-Dimensional Electronic Systems. Annu. Rev. Condens. Matter Phys. 5, 347–373. https://doi.org/10.1146/annurev-conmatphys-031113-133905

Nagai, H. (1974). Structure of vapor-deposited Ga$_x$In$_{1-x}$As crystals. J. Appl. Phys. 45, 3789–3794. https://doi.org/10.1063/1.1663861

Nie, W., Tsai, H., Asadpour, R., Blancon, J.-C., Neukirch, A.J., Gupta, G., Crochet, J.J., Chhowalla, M., Tretiak, S., Alam, M.A., et al. (2015). High-efficiency solution-processed perovskite solar cells with millimeter-scale grains. Science (80-. ). 347, 522–525. https://doi.org/10.1126/science.aaa0472

Oksenberg, E., Merdasa, A., Houben, L., Kaplan-Ashiri, I., Rothman, A., Scheblykin, I.G., Unger, E.L., Joselevich, E. (2020). Large lattice distortions and size-dependent bandgap modulation in epitaxial halide perovskite nanowires. Nat. Commun. 11, 1–11. https://doi.org/10.1038/s41467-020-14365-2

Peedikakkandy, L., Bhargava, P. (2016). Composition dependent optical, structural and photoluminescence characteristics of cesium tin halide perovskites. Rsc Adv. 6, 19857–19860. https://doi.org/10.1039/C5RA2231 7B

Peng, W.C., Wu, Y.S. (2004). High-power AlGaInP light-emitting diodes with metal substrates fabricated by wafer bonding. Appl. Phys. Lett. 84, 1841–1843. https://doi.org/10.1063/1.1682696

Pfeiffer, L., West, K.W., Stormer, H.L., Baldwin, K.W. (1989). Electron mobilities exceeding $10^7$ cm$^2$/V s in modulation-doped GaAs. Appl. Phys. Lett. 55, 1888–1890. https://doi.org/10.1063/1.102162

Pierre, F., Gougam, A.B., Anthore, A., Pothier, H., Esteve, D., Birge, N.O. (2003). Dephasing of electrons in mesoscopic metal wires. Phys. Rev. B 68, 85413. https://doi.org/10.1103/PhysRevB.68.085413

Reyren, N., Thiel, S., Caviglia, A.D., Kourkoutis, L.F., Hammerl, G., Richter, C., Schneider, C.W., Kopp, T., Rüetschi, A.-S., Jaccard, D., et al. (2007). Superconducting Interfaces Between Insulating Oxides. Science (80-. ). 317, 1196–1199. https://doi.org/10.1126/science.1146006

Russell, P.A., Ouali, F.F., Hewett, N.P., Challis, L.J. (1990). Power dissipation in the quantum Hall regime. Surf. Sci. 229, 54–56. https://doi.org/10.1016/0039-6028(90)90831-R

Saidaminov, M.I., Williams, K., Wei, M., Johnston, A., Quintero-Bermudez, R., Vafaie, M., Pina, J.M., Proppe, A.H., Hou, Y., Walters, G., et al. (2020). Multi-cation perovskites prevent carrier reflection from grain surfaces. Nat. Mater. https://doi.org/10.1038/s41563-019-0602-2

Skachko, I., Du, X., Duerr, F., Luican, A., Abanin, D.A., Levitov, L.S., Andrei, E.Y. (2010). Fractional quantum Hall effect in suspended graphene probed with two-terminal measurements. Philos. Trans. R. Soc. A Math. Phys. Eng. Sci. 368, 5403–5416. https://doi.org/10.1098/rsta.2010.0226

Soukiassian, A., Tian, W., Vaithyanathan, V., Haeni, J.H., Chen, L.Q., Xi, X.X., Schlom, D.G., Tenne, D.A., Sun, H.P., Pan, X.Q., et al. (2008). Growth of nanoscale BaTiO$_3$/SrTiO$_3$ superlattices by molecular-beam epitaxy. J. Mater. Res. 23, 1417–1432. https://doi.org/10.1557/JMR.2008.0181

Sun, Y., Thompson, S.E., Nishida, T. (2007). Physics of strain effects in semiconductors and metal-oxide-semiconductor field-effect transistors. J. Appl. Phys. 101, 104503. https://doi.org/10.1063/1.2730561

Sun, Z., Cai, B., Chen, X., Wei, W., Li, X., Yang, D., Meng, C., Wu, Y., Zeng, H. (2020). Prediction and observation of defect-induced room-temperature ferromagnetism in halide perovskites. J. Semicond. 41, 122501. https://doi.org/10.1088/1674-4926/41/12/122501

Takamoto, T., Ikeda, E., Kurita, H., Ohmori, M. (1997). Over 30% efficient InGaP/GaAs tandem solar cells. Appl. Phys. Lett. 70, 381–383. https://doi.org/10.1063/1.118419

Tikhonenko, F.V, Horsell, D.W., Gorbachev, R.V, Savchenko, A.K. (2008). Weak Localization in Graphene Flakes. Phys. Rev. Lett. 100, 56802. https://doi.org/10.1103/PhysRevLett.100.056802

Umansky, V., Heiblum, M., Levinson, Y., Smet, J., Nübler, J., Dolev, M. (2009). MBE growth of ultra-low disorder 2DEG with mobility exceeding 35×10$^6$ cm$^2$/V s. J. Cryst. Growth 311, 1658–1661. https://doi.org/10.1016/j.jcrysgro.2

008.09.151

Wang, L., Chen, P., Kuttipillai, P.S., King, I., Staples, R., Sun, K., Lunt, R.R. (2019). Epitaxial stabilization of tetragonal cesium tin iodide. ACS Appl. Mater. Interfaces 11, 32076–32083. https://doi.org/10.1021/acsami.9b05592

Wang, L., Chen, P., Thongprong, N., Young, M., Kuttipillai, P.S., Jiang, C., Zhang, P., Sun, K., Duxbury, P.M., Lunt, R.R. (2017). Unlocking the Single-Domain Epitaxy of Halide Perovskites. Adv. Mater. Interfaces 4, 1701003. https://doi.org/10.1002/admi.201701003

Wang, L., Meric, I., Huang, P.Y., Gao, Q., Gao, Y., Tran, H., Taniguchi, T., Watanabe, K., Campos, L.M., Muller, D.A., et al. (2013). One-Dimensional Electrical Contact to a Two-Dimensional Material. Science (80-.). 342, 614–617. https://doi.org/10.1126/science.1244358

Wang, Y., Li, X., Song, J., Xiao, L., Zeng, H., Sun, H. (2015). All-Inorganic Colloidal Perovskite Quantum Dots: A New Class of Lasing Materials with Favorable Characteristics. Adv. Mater. 27, 7101–7108. https://doi.org/10.1002/adma.201503573

Wang, Y., Sun, X., Chen, Z., Sun, Y.-Y., Zhang, S., Lu, T.-M., Wertz, E., Shi, J. (2017). High-Temperature Ionic Epitaxy of Halide Perovskite Thin Film and the Hidden Carrier Dynamics. Adv. Mater. 29, 1702643. https://doi.org/10.1002/adma.201702643

Wang, Yiliu, Wan, Z., Qian, Q., Liu, Y., Kang, Z., Fan, Z., Wang, P., Wang, Yekan, Li, C., Jia, C., et al. (2020). Probing photoelectrical transport in lead halide perovskites with van der Waals contacts. Nat. Nanotechnol. 1–8. https://doi.org/10.1038/s41565-020-0729-y

Washburn, S., Webb, R.A. (1986). Aharonov-Bohm effect in normal metal quantum coherence and transport. Adv. Phys. 35, 375–422. https://doi.org/10.1080/00018738600101921

Xu, W., Hu, Q., Bai, S., Bao, C., Miao, Y., Yuan, Z., Borzda, T., Barker, A.J., Tyukalova, E., Hu, Z., et al. (2019). Rational molecular passivation for high-performance perovskite light-emitting diodes. Nat. Photonics 13, 418–424. https://doi.org/10.1038/s41566-019-0390-x

Yang, W.S., Park, B.-W., Jung, E.H., Jeon, N.J., Kim, Y.C., Lee, D.U., Shin, S.S., Seo, J., Kim, E.K., Noh, J.H., et al. (2017). Iodide management in formamidinium-lead-halide based perovskite layers for efficient solar cells. Science (80-.). 356, 1376–1379. https://doi.org/10.1126/science.aan2301

Yuan, M., Quan, L.N., Comin, R., Walters, G., Sabatini, R., Voznyy, O., Hoogland, S., Zhao, Y., Beauregard, E.M., Kanjanaboos, P., et al. (2016). Perovskite energy funnels for efficient light-emitting diodes. Nat. Nanotechnol. 11, 872–877. https://doi.org/10.1038/nnano.2016.110

Yun, Y., Ma, Y., Tao, S., Xing, W., Chen, Y., Su, T., Yuan, W., Wei, J., Lin, X., Niu, Q., et al. (2017). Observation of long phase-coherence length in epitaxial La-doped CdO thin films. Phys. Rev. B 96, 245310. https://doi.org/10.1103/PhysRevB.96.245310

Zhang, L., King, I., Nasyedkin, K., Chen, P., Skinner, B., Lunt, R.R., Pollanen, J. (2021). Coherent Hopping Transport and Giant Negative Magnetoresistance in Epitaxial CsSnBr$_3$. ACS Appl. Electron. Mater. https://doi.org/10.1021/acsaelm.1c00409

# Supplemental Information

## *Extraordinary Phase Coherence Length in Epitaxial Halide Perovskites*

**Kostyantyn Nasyedkin, Isaac King, Liangji Zhang, Pei Chen, Lili Wang, Richard J. Staples, Richard R. Lunt and Johannes Pollanen**

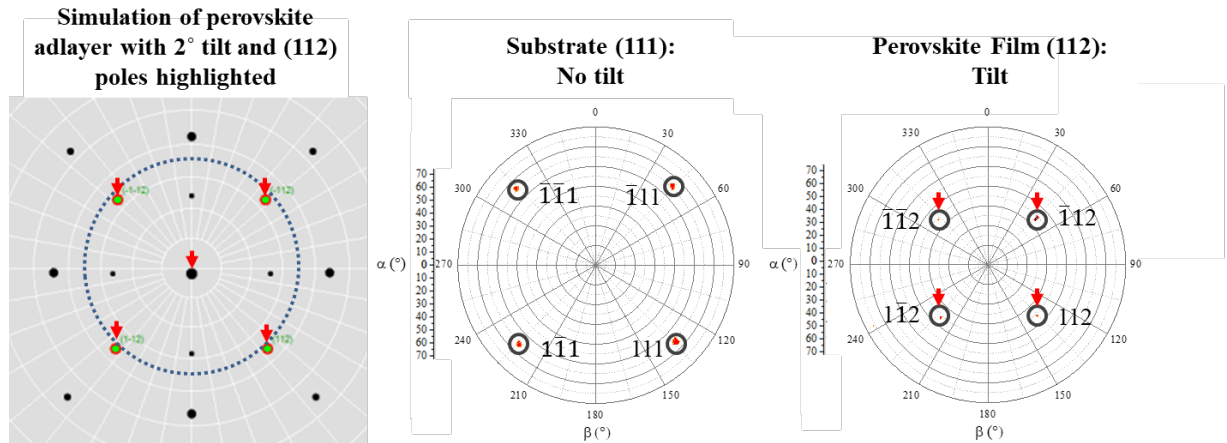

**Figure S1 | Pole figure data and epitaxial tilt.** A simulated pole figure (left) of the CsSnI$_3$ adlayer compared to experimental pole figure scans of the substrate (middle) and perovskite (right), related to Figure 1. The 2% tetragonal distortion in the perovskite adlayer causes the epitaxial layers to grow with a slight tilt to accommodate the pseudomorphic strain at the terrace steps on the substrate, which has a 2% smaller lattice constant. The subsequent pole figure scan then contains poles that are shifted away from their ordinarily radially-symmetric positions, as seen in both the simulated and experimental pole figures. The red arrows indicate direction of the pole shift (lengths exaggerated). This tilt is not observed in the pole figure of the KCl substrate, indicating that the tilt is indeed due to an angled surface of the film (and not from a tilted substrate). This provides further evidence that the films are both epitaxial and pseudomorphic (Nagai, 1974).

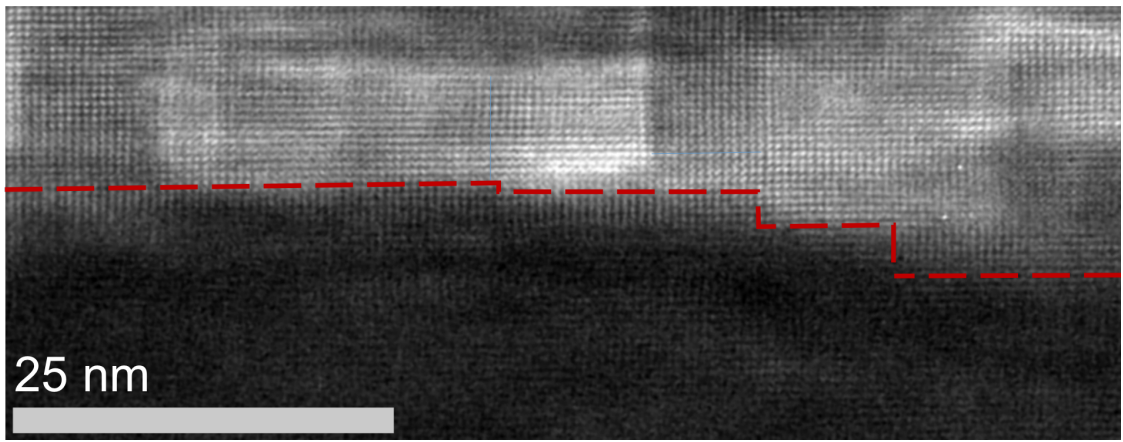

**Figure S2 | TEM image of CsSnI$_3$.** Wide view TEM image shows highly crystalline CsSnI$_3$ adlayer growth upon a terraced surface without clear grain boundaries. Related to Figure 1. The dashed redline indicates the interface between the CsSnI$_3$ epilayer and the KCl substrate. We note that the areas of differing brightness likely arise from local variations in the thickness of the TEM slice.

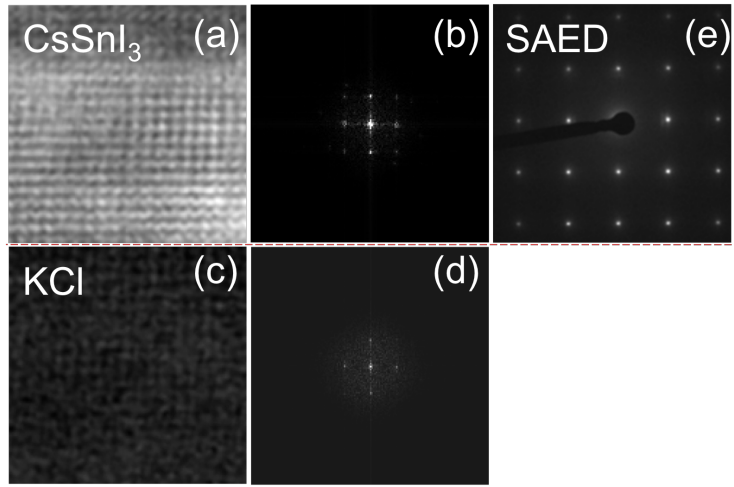

**Figure S3 | TEM analysis and selected area electron diffraction of CsSnI₃ epilayer.** TEM images and associated Fast Fourier Transform (FFT) analysis for the CsSnI₃ adlayer (panels a, b) and KCl substrate (panels c, d) demonstrating the epitaxial relation between the adlayer and substrate. Related to Figure 1. We note that tetragonal distortion of the CsSnI3 layer should be observed in the FFT images but is difficult to see from the pattern due to the small amount of tetragonal distortion (< 3%). In fact the tetragonal distortion is more clearly be extracted from the TEM image by counting rows of atoms horizontally versus vertically. Selected area electron diffraction (SAED) for the CsSnI₃ layer shown in panel (e) clearly shows the tetragonal distortion that leads to the tensile reduction in the crystal c-direction.

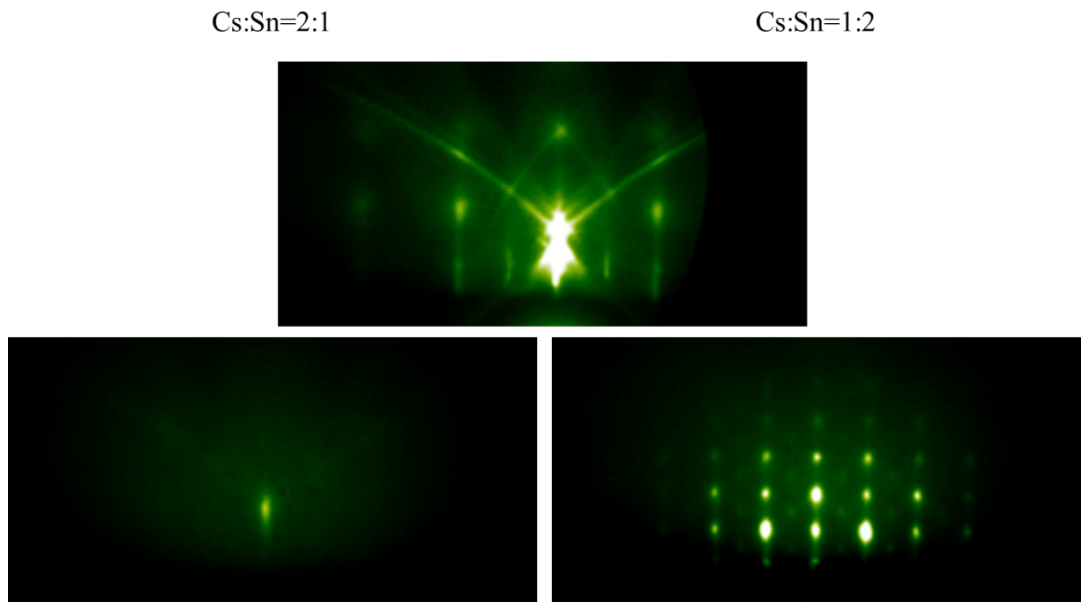

**Figure S4 | Off-stoichiometric growth characterization.** RHEED scans of CsSnI₃ on KCl with varying precursor ratios of CsI:SnI₂ showing changes in crystal structure and quality as growth leaves the stoichiometric ratio of 1:1. Related to Figure 1. The scans show the base substrate (top) and the epitaxial film at several hundred angstroms (bottom). The 2:1 growth (Sn vacancies), results in an essentially amorphous film (confirmed via ex-situ XRD), and can be seen in the complete lack of clear streaks or spots in the RHEED pattern. The 1:2 growth (Cs vacancies), results in a rough crystalline film that is likely CsSn₂I₅, analogous to a similar bromide (CsSn₂Br₅) compound seen in our previous work (L. Wang et al., 2017).

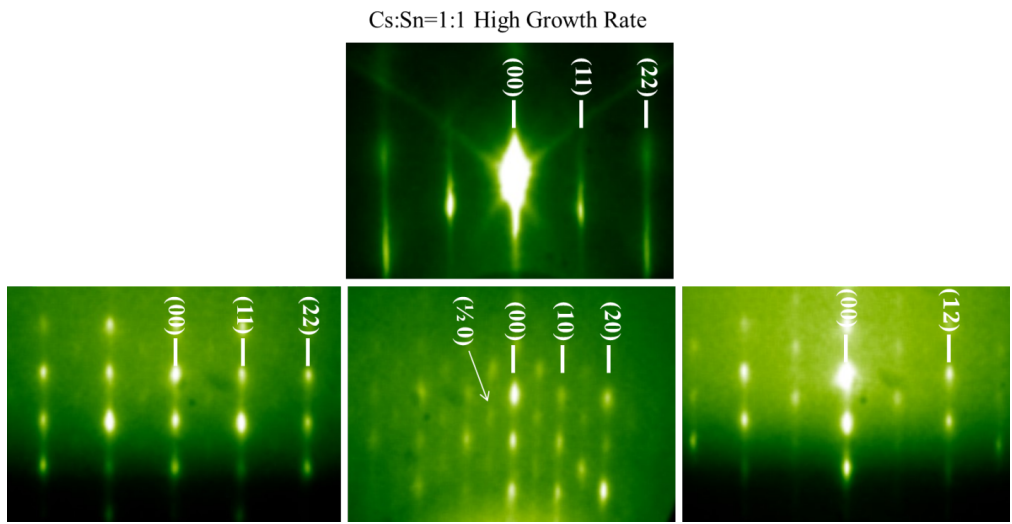

**Figure S5 | Growth rate characterization.** RHEED scans of KCl (top image) with the CsSnI$_3$ growth (bottom images) performed at a high growth rate (~0.1 nm/s). The bottom images are different rotations of the final film, showing spotty (rough) patterns. The low growth quality at high rates likely stems from imbalance of the lower reaction rate with high deposition rate that leads to an accumulation of vacancies and non-uniform (island) growth. The presence of half order spots indicates a larger unit cell that may result from the presence of Cs vacancies similar to the 1:2 growth shown in Figure S4 above. Related to Figure 1.

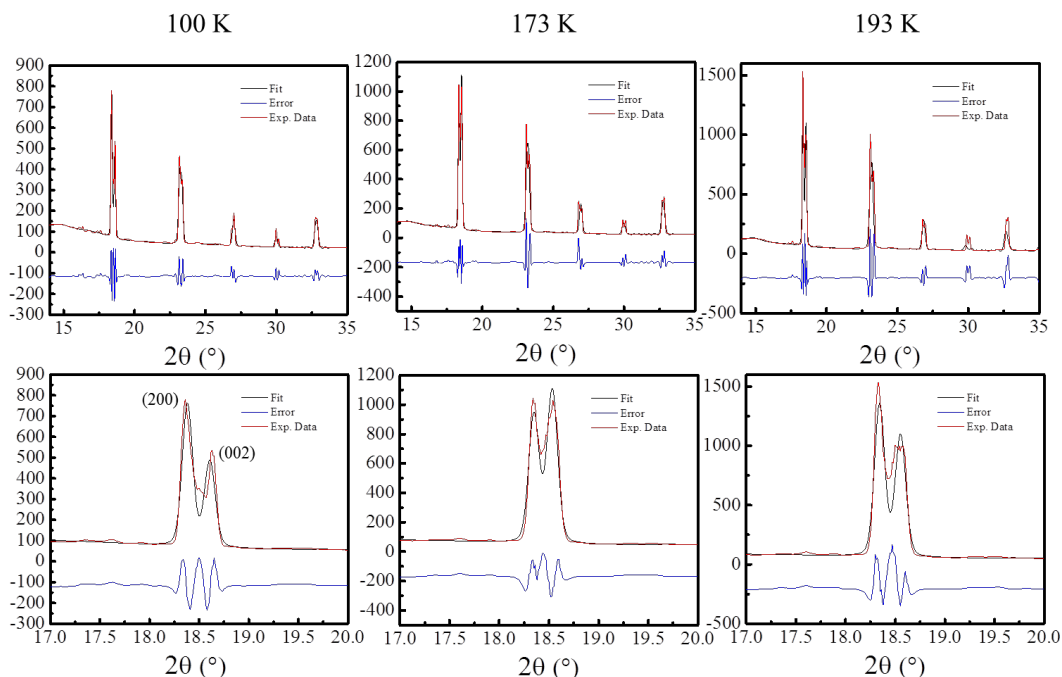

**Figure S6 | Synchrotron characterization at low temperature.** Refinement (Pawley) fits of experimental synchrotron data made using the Reflex module in Materials Studio ($\lambda$ = 0.61987Å). Related to Figure 1. Note that the sample was formed from folding an epitaxial film that was lifted-off and thus is not a perfect powder but rather a strongly textured sample. Top panel shows fits of the full patterns and bottom shows a zoom in on the (002)/(200) peak splitting indicative of the tetragonal structure at all temperatures. Extracted lattice constants are provided in Table S1 below.

| T (K) | | Å | R |
|---|---|---|---|
| 100 | a | 6.198 | 1.02 |
| | c | 6.075 | |
| 173 | a | 6.205 | 1.02 |
| | c | 6.107 | |
| 193 | a | 6.214 | 1.02 |
| | c | 6.107 | |

**Table S1 | Synchrotron characterization results.** Lattice constants and a/c ratios extracted from fittings of experimental XRD synchrotron data at three different temperatures. Related to Figure 1

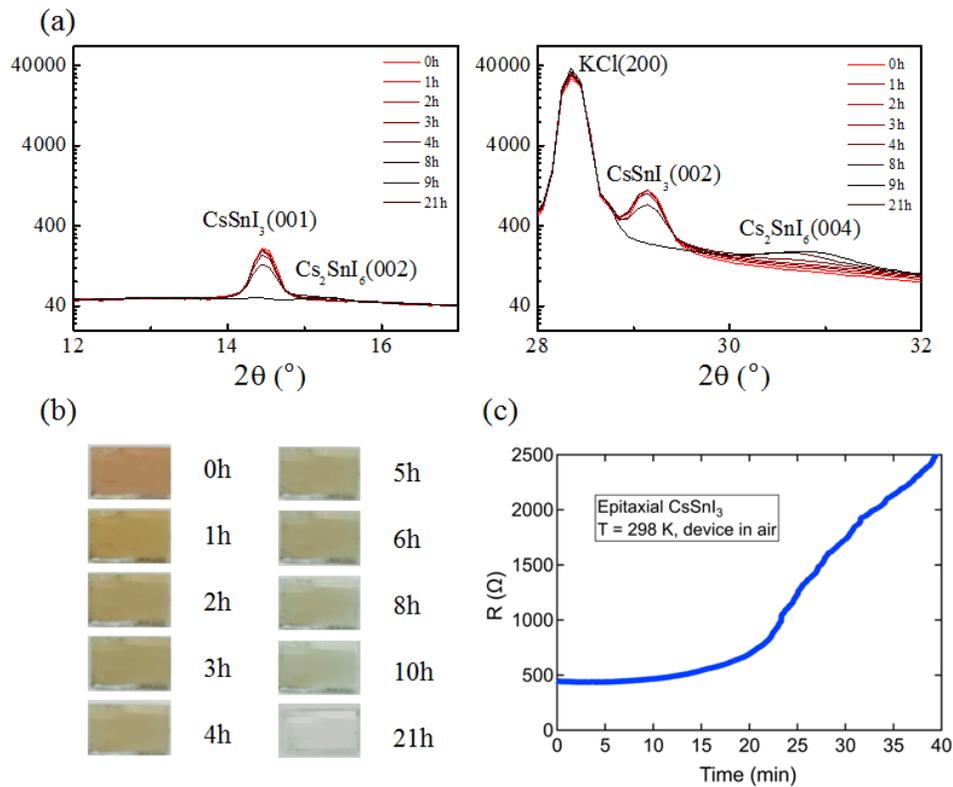

**Figure S7 | Characterization of air sensitivity.** (a) XRD scans of a 50 nm epitaxial $CsSnI_3$ film degrading over time while exposed to air. At 3 hours, the perovskite peaks begin to disappear and the degradation product, $Cs_2SnI_6$, appears. (b) Optical images of the degrading perovskite sample, showing the transformation from the brown $CsSnI_3$ to the transparent $Cs_2SnI_6$. (c) Resistance over time of a $CsSnI_3$ film device exposed to air. Changes in the resistance begin to occur after ~ 20 min (faster than in XRD), likely due to the degradation at the interface first and a greater sensitivity in the resistance measurement overall. After several hours the resistance becomes > 50MΩ. Related to STAR Method, Method Details, $CsSnI_3$ sample growth.

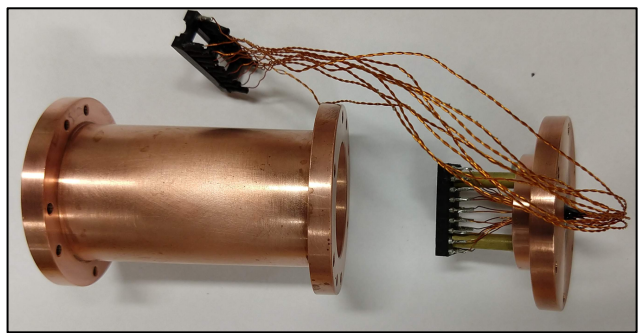

**Figure S8 | Hermetically sealed sample cell.** The devices are loaded and sealed within the sample cell in a dry nitrogen environment. Related to STAR Method, Method Details, $CsSnI_3$ sample growth.

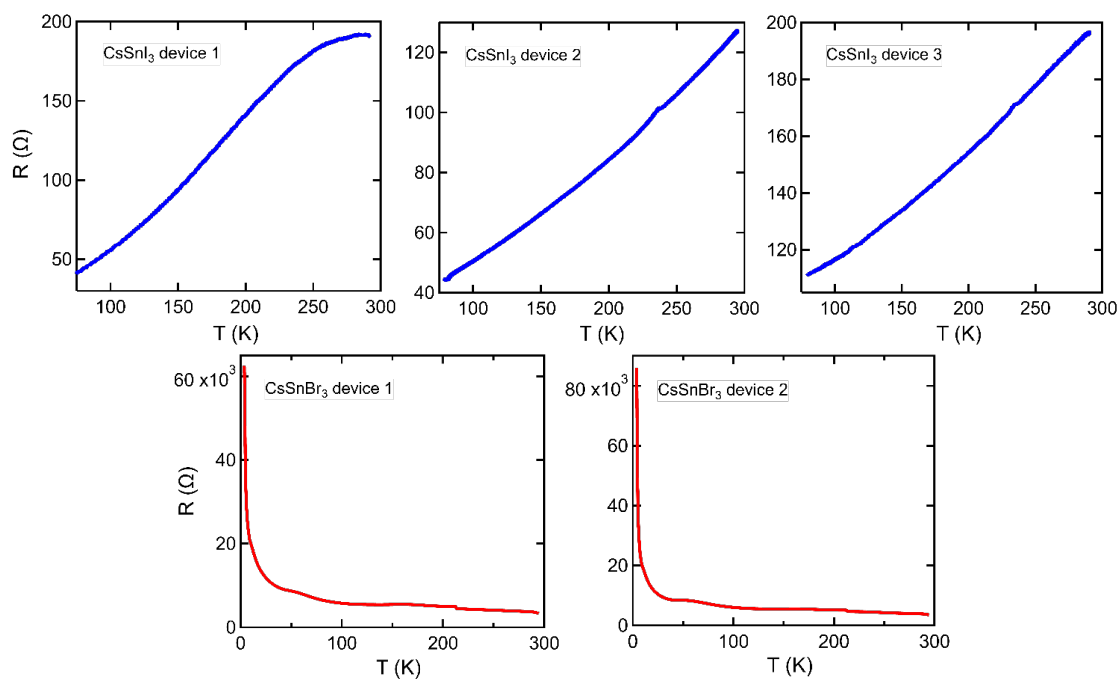

**Figure S9 | Resistance $R$ versus temperature $T$ of different $CsSnI_3$ and $CsSnBr_3$ devices.** Epitaxial $CsSnI_3$ devices compared to epitaxial $CsSnBr_3$ devices. Related to Figure 2. Both of these epitaxial $CsSnBr_3$ devices show an increasing resistance with decreasing temperature, which is markedly distinct from $CsSnI_3$ and serves as a control to demonstrate that the behavior described in the paper is arising from the $CsSnI_3$ epilayer.

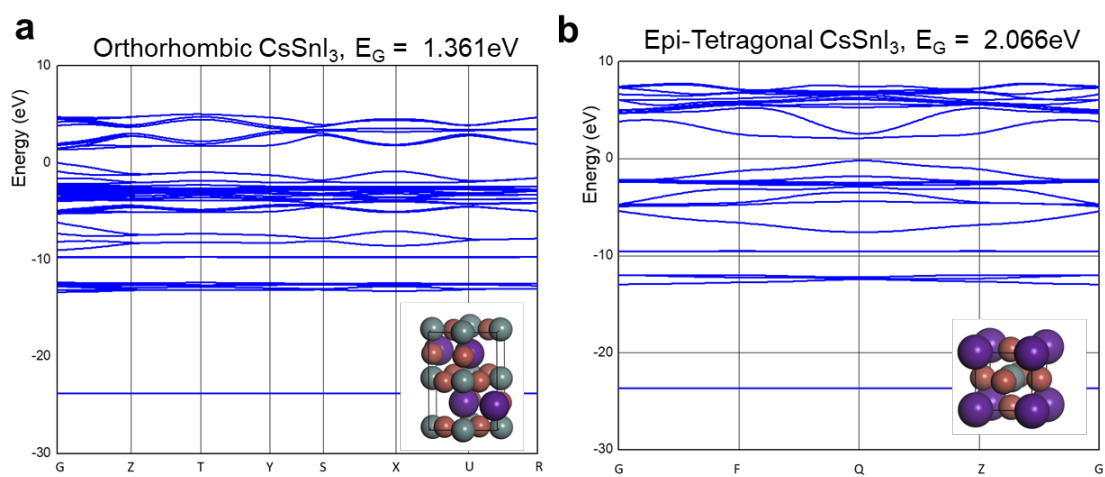

**Figure S10 | Calculated electronic band structures.** Calculated band structures for the (a) orthorhombic and (b) epi-tetragonal phases of $CsSnI_3$. Related to Figure 2. The data were calculated in Materials Studio 7.0 using the B3LYP DFT functional in the CASTEP module.

| Phase | Cubic (High $T$) | Tetragonal (High $T$) | Orthorhombic (Room $T$) | Epi-tetragonal (Room $T$, this work) |
|---|---|---|---|---|
| **Crystal Structure** <br> ● Cs <br> ● Sn <br> ● I | | | | |
| **Lattice Parameters (nm)** | $a$=0.6206 | $a$=$b$=0.8718, $c$=0.6191 | $a$=0.8689, $b$=1.2378, $c$=0.8638 | $a$=$b$=0.622±0.007, $c$=0.612±0.002 |
| **Experimental Bandgap (eV)** | - | - | 1.31 (Peedikakkandy and Bhargava, 2016) | 1.85 (L. Wang et al., 2017) |
| **Simulated Bandgap (eV)** | 0.76 | 0.41 | 1.40 | 2.07 |

**Table S2 | Crystallographic Data of Various Phases.** Modified from (Groom et al., 2016; Wang et al., 2019). Simulated bandgaps were calculated in Materials Studio 7.0 using the CASTEP module with the B3LYP functional. Related to Figure 2. At room temperature, only the orthorhombic and epi-tetragonal phases have been observed for $CsSnI_3$, along with a yellow (large bandgap) orthorhombic phase (not included).